# Positive Definite Estimation of Large Covariance Matrix Using Generalized Nonconvex Penalties

Fei Wen, *Member, IEEE,* Yuan Yang, Peilin Liu, *Member, IEEE*, Robert C. Qiu, *Fellow, IEEE*

*Abstract*—This work addresses the issue of large covariance matrix estimation in high-dimensional statistical analysis. Recently, improved iterative algorithms with positive-definite guarantee have been developed. However, these algorithms cannot be directly extended to use a nonconvex penalty for sparsity inducing. Generally, a nonconvex penalty has the capability of ameliorating the bias problem of the popular convex lasso penalty, and thus is more advantageous. In this work, we propose a class of positive-definite covariance estimators using generalized nonconvex penalties. We develop a first-order algorithm based on the alternating direction method framework to solve the nonconvex optimization problem efficiently. The convergence of this algorithm has been proved. Further, the statistical properties of the new estimators have been analyzed for generalized nonconvex penalties. Moreover, extension of this algorithm to covariance estimation from sketched measurements has been considered. The performances of the new estimators have been demonstrated by both a simulation study and a gene clustering example for tumor tissues. Code for the proposed estimators is available at https://github.com/FWen/Nonconvex-PDLCE.git.

## I. INTRODUCTION

Nowadays, the advance of information technology makes massive high-dimensional data widely available for scientific discovery, which makes Big Data a very hot research topic. In this context, effective statistical analysis for high-dimensional data is becoming increasingly important. In much statistical analysis of high-dimensional data, estimating large covariance matrices is needed, which has attracted significant research attentions in the past decade and has found applications in many fields, such as economics and finance, bioinformatics, social networks, smart grid, and climate studies [1-5]. In these applications, the covariance information is necessary for effective dimensionality reduction and discriminant analysis. The goal of covariance estimation is to recover the population covariance matrix of a distribution from independent and identically distributed samples. In the high-dimensional setting, the dimensionality is often comparable to (or even larger than) the sample size, in which cases the standard sample covariance matrix estimator has a poor performance, since the number of unknown parameters grows quadratically in the dimension [2, 6, 7].

To achieve better estimation of large covariance matrix, intrinsic structures of the covariance matrix can be exploited by using regularization techniques, such as banding and tapering for banded structure [8-12], thresholding for sparse structure [13-15]. The former is useful for the applications where the variables have a natural ordering and variables far apart are only weakly correlated, such as in longitudinal data, time series, spatial data, or spectroscopy. For other applications, where the variables do not have such properties but the true covariance matrix is sparse, permutation-invariant thresholding methods have been proposed in [13, 14]. These methods have good theoretical properties and are computationally efficient. It has been shown in [15] that the generalized thresholding estimators are consistent over a large class of (approximately) sparse covariance matrices. However, in practical finite sample applications, such an estimator is not always positive-definite although it converges to a positive-definite limit in the asymptotic setting.

Positive definiteness is desirable in many statistical learning applications such as quadratic discriminant analysis and covariance regularized regression [16]. To simultaneously achieve sparsity and positive definiteness, iterative methods have been proposed recently in [17-20]. In [17], a positive-definite estimator has been proposed via maximizing a penalized Gaussian likelihood with a lasso penalty, and a majorize-minimize algorithm has been designed to solve the estimation problem. In [18], a logarithmic barrier term is added into the objective function of the soft-thresholding estimator to enforce positive-definiteness. Then, new positive-definite estimators have been proposed in [19, 20] by imposing an eigenvalue constraint on the optimization problem of the soft-thresholding estimator. Although these positive-definite estimators have good theoretical properties, the derived algorithms are restricted to the convex $\ell_1$-norm (lasso) penalty and cannot be directly extended to use a nonconvex penalty, since they are not guaranteed to converge in that case. In [20], in addition to the $\ell_1$-penalty, the nonconvex minimax concave

This work was supported in part by the National Natural Science Foundation of China (NSFC) under grants 61401501, 61573242 and 61472442, and by the Young Star Science and Technology Project in Shaanxi province under grant 2015KJXX-46.
F. Wen, P. Liu and R. C. Qiu are with the Department of Electronic Engineering, Shanghai Jiao Tong University, Shanghai 200240, China (e-mail: wenfei@sjtu.edu.cn; liupeilin@sjtu.edu.cn; rcqiu@sjtu.edu.cn).
F. Wen is also with the Air Control and Navigation Institution, Air Force Engineering University, Xian 710000, China.
Y. Yang is with the Air Control and Navigation Institution, Air Force Engineering University, Xian 710000, China.

(MC) penalty has also been considered and an algorithm has been proposed based on local linear approximation of the MC penalty.

Compared with the convex $\ell_1$-penalty, a nonconvex penalty, such as the hard-thresholding or smoothly clipped absolute deviation (SCAD), is more advantageous since it can ameliorate the bias problem of the $\ell_1$-one. This work proposes a class of positive-definite covariance estimators using generalized nonconvex penalties. We use an eigenvalue constraint to ensure the positive definiteness of the estimator similar to [19, 20], but the penalty can be noncovex. With an eigenvalue constraint of the covariance and simultaneously employing a nonconvex penalty make the optimization problem challenging. To solve the nonconvex optimization problem efficiently, we present a first-order algorithm based on the alternating direction method (ADM) framework. It has been proved that the sequence generated by the proposed algorithm converges to a stationary point of the objective function if the penalty is a Kurdyka-Lojasiewicz (KL) function. Further, the statistical properties of the new estimator have been analyzed for a generalized nonconvex penalty. The effectiveness of the new estimators has been demonstrated via both a simulation study and a real gene clustering experiment.

Moreover, extension of the proposed ADM algorithm to sparse covariance estimation from *sketches* or *compressed measurements* has also been considered. Covariance sketching is an efficient approach for covariance estimation from high-dimensional data stream [36-42]. In many practical applications to extract the covariance information from high-dimensional data stream at a high rate, it may be infeasible to sample and store the whole stream due to memory and processing power constraint. In this case, by exploiting the structure information of the covariance matrix, it can be reliably recovered from compressed measurements of the data stream with a significantly lower dimensionality. The proposed algorithm can simultaneously achieve positive-definiteness and sparsity in estimating the covariance from compressed measurements.

The rest of this paper is organized as follows. Section II introduces the background of large covariance estimation. In section III, we detail the new algorithm and present some analysis on its convergence property. Section IV contains the statistical properties of the new method. In section V, we extend the new method to positive-definite covariance estimation from compressed measurements. Section VI contains experimental results. Finally, Section VII ends the paper with concluding remarks.

The following notations are use throughout the paper. For a matrix $\mathbf{M}$, $\mathrm{diag}(\mathbf{M})$ is a diagonal matrix which has the same diagonal elements as that of $\mathbf{M}$, whilst for a vector $\mathbf{v}$, $\mathrm{diag}(\mathbf{v})$ is a diagonal matrix with diagonal elements be $\mathbf{v}$. $I(\cdot)$ denotes the indicator function. $\mathbf{I}_m$ stands for an $m\times m$ identity matrix and $\mathbf{1}_{m\times m}$ denotes an $m\times m$ matrix with all elements be one. $(\cdot)^T$ denotes the transpose operator. $\varphi_{\min}(\mathbf{M})$ and $\varphi_{\max}(\mathbf{M})$ stand for the minimum and maximal eigenvalues of $\mathbf{M}$, respectively. $\circ$, $\oslash$ and $\otimes$ stand for the Hadamard product, Hadamard division and Kronecker product, respectively. $\mathrm{vec}(\mathbf{M})$ is the "vectorization" operator stacking the columns of the matrix one below another. $\mathrm{dist}(\mathbf{X},S):=\inf\{\|\mathbf{Y}-\mathbf{X}\|_F:\mathbf{Y}\in S\}$ denotes the distance from a point $\mathbf{X}\in\mathbb{R}^{m\times n}$ to a subset $S\subset\mathbb{R}^{m\times n}$. $\mathbf{X}\geq 0$ and $\mathbf{X}>0$ mean that $\mathbf{X}$ is positive-semidefinite and positive-definite, respectively.

## II. BACKGROUND

For a vector $\mathbf{x}\in\mathbb{R}^d$ with covariance $\mathbf{\Sigma}^0=E\{\mathbf{x}\mathbf{x}^T\}$, the goal is to estimate the covariance from $n$ observations $\mathbf{x}_1,\cdots,\mathbf{x}_n$. In this work, we are interested in estimating the correlation matrix $\mathbf{\Theta}^0=\mathrm{diag}(\mathbf{\Sigma}^0)^{-1/2}\mathbf{\Sigma}^0\mathrm{diag}(\mathbf{\Sigma}^0)^{-1/2}$, where $\mathbf{\Theta}^0$ is the true correlation matrix, and $\mathrm{diag}(\mathbf{\Sigma}^0)^{-1/2}$ is the diagonal matrix of true standard deviations. With the estimated correlation matrix, denoted by $\hat{\mathbf{\Theta}}$, the estimation of the covariance matrix is $\hat{\mathbf{\Sigma}}=\mathrm{diag}(\mathbf{R})^{1/2}\hat{\mathbf{\Theta}}\mathrm{diag}(\mathbf{R})^{1/2}$, where $\mathbf{R}$ denotes the sample covariance matrix. This procedure is more favorable than that of estimating the covariance matrix directly, since the correlation matrix retains the same sparsity structure of the covariance matrix but with all the diagonal elements known to be one. Since the diagonal elements need not to be estimated, the correlation matrix can be estimated more accurately than the covariance matrix [20-22].

*A. Generalized Thresholding Estimator*

Given the sample correlation matrix $\mathbf{S}$, the generalized thresholding estimator [15] solves the following problem

$$\min_{\mathbf{\Theta}}\frac{1}{2}\|\mathbf{\Theta}-\mathbf{S}\|_F^2+g_\lambda(\mathbf{\Theta}) \qquad (1)$$

where $g_\lambda(\mathbf{\Theta})$ is a generalized penalty function depending on a penalty parameter $\lambda$. The penalty function can be expressed in an element-wise form as $g_\lambda(\mathbf{\Theta})=\sum_{i\neq j}g_\lambda(\mathbf{\Theta}_{ij})$, which only penalizes the off-diagonal elements since the diagonal elements of a correlation (also covariance) matrix are always positive. The thresholding operation corresponding to $g_\lambda(\cdot)$ is defined as

$$T_\lambda(x)=\min_z\left\{\frac{1}{2}(z-x)^2+g_\lambda(z)\right\}. \qquad (2)$$

For the popular SCAD, hard-, $\ell_q$-, and soft-thresholding, the penalties and corresponding thresholding functions are given as follows.

(i) Hard-thresholding. The penalty is given by [23]

$$g_\lambda(x)=\lambda^2-(|x|-\lambda)^2 I(|x|<\lambda)$$

and the corresponding thresholding function is

$$T_\lambda(x) = xI(|x|>\lambda). \quad (3)$$

Note that, the $\ell_0$-norm penalty

$$g_\lambda(x) = \frac{\lambda^2}{2}|x|_0 = \frac{\lambda^2}{2}I(|x|\neq 0)$$

also results in (3).

(ii) Soft-thresholding, $g_\lambda(x)=\lambda|x|$. The corresponding thresholding function is [24]

$$T_\lambda(x) = \text{sign}(x)(|x|-\lambda)_+.$$

Soft-thresholding is widely used in sparse variable selection, since the $\ell_1$-norm minimization problem is more tractable (due to its convexity) than the problems using nonconvex penalties. However, the $\ell_1$-penalty has a bias problem as lasso soft-thresholding would produce biased estimates for large coefficients. This bias problem can be ameliorated by using a nonconvex penalty, e.g., Hard-thresholding or SCAD.

(iii) SCAD. The penalty is given by

$$p_\lambda(x) = \begin{cases} \lambda|x|, & 0\leq|x|<\lambda \\ (2a\lambda|x|-x^2-\lambda^2)/[2(a-1)], & \lambda\leq|x|<a\lambda \\ (a+1)\lambda^2/2, & |x|\geq a\lambda \end{cases}$$

for some $a>2$. The corresponding thresholding function is [25]

$$T_\lambda(x) = \begin{cases} \text{sign}(x)(|x|-\lambda)_+, & |x|\leq 2\lambda \\ [(a-1)x-\text{sign}(x)a\lambda]/(a-2), & 2\lambda<|x|\leq a\lambda \\ x, & |x|>a\lambda \end{cases}.$$

(iv) $\ell_q$-norm ($0<q<1$), $p_\lambda(x)=\alpha(\lambda,q)|x|^q$ where $\alpha(\lambda,q)=(\lambda-\beta)\beta^{1-q}/q$ with $\beta=2(1-q)\lambda/(2-q)$. The thresholding function is [26, 31]

$$T_\lambda(x) = \begin{cases} 0, & |x|<\lambda \\ \{0,\text{sign}(x)\beta\}, & |x|=\lambda \\ \text{sign}(x)\eta, & |x|>\lambda \end{cases} \quad (4)$$

where $\eta$ is the solution of $h(\eta)=\alpha q\eta^{q-1}+\eta-x=0$ over the region $(\beta,|x|)$. Since $h(\eta)$ is convex, when $|x|>\lambda$, $\eta$ can be efficiently solved using a Newton's method, e.g.,

$$\eta^{k+1} = \eta^k - \frac{h(\eta^k)}{h'(\eta^k)}, \quad k=0,1,2,\cdots.$$

The starting point can be simply chosen as $\eta^0=|x|$. For the special cases of $q=1/2$ or $q=2/3$, the proximal mapping can be explicitly expressed as the solution of a cubic or quartic equation [27].

Fig. 1 shows the generalized thesholding\shrinkage functions for the hard-, soft-, $\ell_q$- and SCAD penalties with a same threshold. When the true parameter has a relatively large magnitude, the soft-thresholding estimator is biased since it imposes a constant shrinkage on the parameter. In contrast, the hard-thresholding and SCAD estimators are unbiased for large parameter. Another idea to mitigate the bias of soft-thresholding is to employ an adaptive penalty [28]. The thresholding rule corresponding to this adaptive penalty, as well as SCAD and $\ell_q$, fall in (sandwiched) between hard- and soft-thresholding.

These generalized covariance estimators are computationally efficient, e.g., the SCAD, hard- and soft-thresholding estimators only need an element-wise thresholding operation of the sample covariance matrix. For the adaptive lasso estimator [28], only a few iterations are sufficient to achieve satisfactory performance, and each iteration only need a soft-thresholding operation. However, these estimators are not always positive definite in practical finite sample applications, although the positive definiteness can be guaranteed in the asymptotic setting with probability tending to 1. Intuitively, to deal with the infiniteness problem, we can project the estimate $\hat{\Theta}$ into a convex cone $\{\Theta\geq 0\}$. Specifically, let $\hat{\Theta}=\sum_{i=1}^d \lambda_i \mathbf{v}_i^T \mathbf{v}_i$ denote the eigen-decomposition of $\hat{\Theta}$, where $\lambda_i$ is eigenvalue corresponding to the eigenvector $\mathbf{v}_i$. A positive-semidefinite estimate can be obtained as $\hat{\Theta}^+ = \sum_{i=1}^d \max(\lambda_i,0)\mathbf{v}_i^T \mathbf{v}_i$. However, this projection would destroy the sparsity pattern of the true correlation matrix and result in a non-sparse estimate (see [19] for a detailed example).

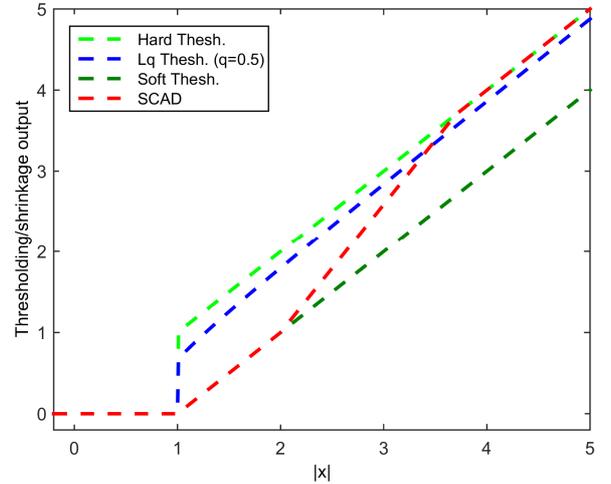

Fig. 1. Generalized thesholding\shrinkage functions for the hard-, soft-, $\ell_q$- and SCAD penalties.

*B. Positive Definite Estimator*

To simultaneously achieve sparsity and positive-definiteness, iterative methods have been proposed recently in [18-20], e.g., the constrained correlation matrix estimator which solves the following problem [20]

$$\min_\Theta \frac{1}{2}\|\Theta-\mathbf{S}\|_F^2 + \lambda\|\Theta\|_{1,\text{off}}$$

$$\text{subject to} \quad \text{diag}(\Theta)=\mathbf{I}_d \text{ and } \Theta\geq\varepsilon_1\mathbf{I}_d \quad (5)$$

where $\varepsilon_1 > 0$ is the lower bound for the minimum eigenvalue. A covariance matrix version of (5) has been proposed in [19]. Unlike the works [19] and [20] imposing an eigenvalue constraint, the method in [18] employs a logarithmic barrier term to ensure the positive-definiteness of the solution. These works focus mainly on the $\ell_1$-norm penalty, since it results in convex optimization problems which can be efficiently solved with convergence guarantee. In [20], in addition to the $\ell_1$-norm penalty, the nonconvex minimax concave penalty has also been considered, which retains the global convexity of the problem when a tuning parameter is appropriately selected. However, as mentioned by the authors, the derived augmented Lagrangian method (ALM) based algorithm often fails to converge. To address this problem, an alternative algorithm using local linear approximation has been proposed in [20].

Generally, these iterative algorithms cannot be directly extended to use a nonconvex penalty, since they are not guaranteed to converge in that case. With a nonconvex penalty, the optimization problem is more difficult to handle than the convex case.

III. PROPOSED ESTIMATORS USING GENERALIZED NONCONVEX PENALTIES

In this section, we propose a class of positive-definite covariance estimator using generalized penalties as follows

$$\min_{\boldsymbol{\Theta}} \frac{1}{2}\|\boldsymbol{\Theta} - \mathbf{S}\|_F^2 + g_\lambda(\boldsymbol{\Theta})$$
$$\text{subject to} \quad \text{diag}(\boldsymbol{\Theta}) = \mathbf{I}_d \text{ and } \boldsymbol{\Theta} \geq \varepsilon_1 \mathbf{I}_d \quad (6)$$

where $g_\lambda$ a generalized (may be nonconvex) penalty, such as SCAD, hard-, soft-, or $\ell_q$-penalty. The problem (6) can be equivalently rewritten as

$$\min_{\boldsymbol{\Theta}} \frac{1}{2}\|\boldsymbol{\Theta} - \mathbf{S}\|_F^2 + g_\lambda(\boldsymbol{\Theta}) + \delta_{\Omega_1}(\boldsymbol{\Theta})$$
$$\text{subject to} \quad \text{diag}(\boldsymbol{\Theta}) = \mathbf{I}_d \quad (7)$$

where $\Omega_1 := \{\boldsymbol{\Theta} : \boldsymbol{\Theta} \geq \varepsilon_1 \mathbf{I}_d\}$ is a closed convex set, $\delta_{\Omega_1}(\cdot)$ denotes the indicator function defined on this set, i.e.,

$$\delta_{\Omega_1}(\boldsymbol{\Theta}) = \begin{cases} 0, & \boldsymbol{\Theta} \in \Omega_1 \\ +\infty, & \text{otherwise} \end{cases}.$$

The problem (7) is generally difficult to solve since in addition to the nonconvexity of the penalty term $g_\lambda$, both the terms $g_\lambda$ and $\delta_{\Omega_1}$ are nonsmooth. For nonconvex $g_\lambda$, the ALM algorithms proposed in [19, 20] cannot be directly used to solve (7), since these algorithms may fail to converge in this case. In the following, we propose an ADM algorithm to solve the problem (7).

ADM is a powerful optimization framework that is well suited to large-scale problems arising in machine learning and signal processing. In the ADM framework, the three terms in (7) are naturally separated, which makes the problem easy to tackle as each step of the alternating minimization is much easier than the global problem. More specifically, using two auxiliary variables $\mathbf{V}_1, \mathbf{V}_2 \in \mathbb{R}^{d \times d}$, the problem (7) can be equivalently rewritten as

$$\min_{\boldsymbol{\Theta}, \mathbf{V}_1, \mathbf{V}_2} \frac{1}{2}\|\boldsymbol{\Theta} - \mathbf{S}\|_F^2 + g_\lambda(\mathbf{V}_1) + \delta_{\Omega_1}(\mathbf{V}_2)$$
$$\text{subject to} \quad \text{diag}(\mathbf{V}_1) = \mathbf{I}_d, \mathbf{G}\boldsymbol{\Theta} = \mathbf{V} \quad (8)$$

with

$$\mathbf{V} = \begin{bmatrix} \mathbf{V}_1 \\ \mathbf{V}_2 \end{bmatrix}, \quad \mathbf{G} = \begin{bmatrix} \mathbf{I}_d \\ \mathbf{I}_d \end{bmatrix}.$$

The constrained minimization problem (8) can be attacked via solving

$$\min_{\boldsymbol{\Theta}, \mathbf{V}_1, \mathbf{V}_2} \frac{1}{2}\|\boldsymbol{\Theta} - \mathbf{S}\|_F^2 + g_\lambda(\mathbf{V}_1) + \delta_{\Omega_1}(\mathbf{V}_2) + \frac{\rho}{2}\|\mathbf{G}\boldsymbol{\Theta} - \mathbf{V}\|_F^2$$
$$\text{subject to} \quad \text{diag}(\mathbf{V}_1) = \mathbf{I}_d \quad (9)$$

where $\rho > 0$ is a penalty parameter. For sufficiently large $\rho$, e.g., $\rho \to +\infty$, the solution of (9) approaches that of the problem (8). In practical applications, selecting a moderate value of $\rho$ suffices to achieve satisfactory performance. Then, ADM applied to (9) consists of the following three steps in the $k+1$-th iteration

$$\mathbf{V}_1^{k+1} = \arg\min_{\text{diag}(\mathbf{V}_1) = \mathbf{I}_d} \left( g_\lambda(\mathbf{V}_1) + \frac{\rho}{2}\|\boldsymbol{\Theta}^k - \mathbf{V}_1\|_F^2 + \frac{c_k}{2}\|\mathbf{V}_1 - \mathbf{V}_1^k\|_F^2 \right)$$
(10)

$$\mathbf{V}_2^{k+1} = \arg\min_{\mathbf{V}_2} \left( \delta_{\Omega_1}(\mathbf{V}_2) + \frac{\rho}{2}\|\boldsymbol{\Theta}^k - \mathbf{V}_2\|_F^2 + \frac{d_k}{2}\|\mathbf{V}_2 - \mathbf{V}_2^k\|_F^2 \right) \quad (11)$$

$$\boldsymbol{\Theta}^{k+1} = \arg\min_{\boldsymbol{\Theta}} \left( \frac{1}{2}\|\boldsymbol{\Theta} - \mathbf{S}\|_F^2 + \frac{\rho}{2}\|\mathbf{G}\boldsymbol{\Theta} - \mathbf{V}^{k+1}\|_F^2 \right) \quad (12)$$

where $c_k > 0$ and $d_k > 0$. This method considers the proximal regularization of the Gauss-Seidel scheme via coupling the Gauss-Seidel iteration scheme with a proximal term. Using this proximal regularization strategy, as will be shown later, the sequence generated via (10)-(12) is guaranteed to converge in the case of a nonconvex $g_\lambda$.

The $\mathbf{V}_1$-subproblem (10) is a proximal minimization problem, which has a solution as

$$(\mathbf{V}_1^{k+1})_{ij} = \begin{cases} T_{\lambda/(\rho+c_k)} \left( \frac{\rho \boldsymbol{\Theta}_{ij}^k + c_k(\mathbf{V}_1^k)_{ij}}{\rho + c_k} \right), & i \neq j \\ 1, & i = j \end{cases} \quad (13)$$

where $T_\alpha(\cdot)$ is the thresholding operation corresponding to the penalty function $g_\lambda$. The $\mathbf{V}_2$-subproblem (11) is also a proximal minimization problem, whose solution is given by

$$\mathbf{V}_2^{k+1} = \mathrm{P}_+\left(\frac{\rho\boldsymbol{\Theta}^k + d_k \mathbf{V}_2^k}{\rho + d_k}, \varepsilon_1\right). \quad (14)$$

where $\mathrm{P}_+(\cdot,\cdot)$ is a spectral projection operator. Let $\mathbf{M} = \sum_{i=1}^d \lambda_i \mathbf{v}_i^T \mathbf{v}_i$ denote the eigen-decomposition of $\mathbf{M}$, where $\lambda_i$ and $\mathbf{v}_i$ are the eigenvalues and eigenvectors, $\mathrm{P}_+(\cdot,\cdot)$ is defined as $\mathrm{P}_+(\mathbf{M},\varepsilon_1) = \sum_{i=1}^d \max(\lambda_i, \varepsilon_1)\mathbf{v}_i^T \mathbf{v}_i$.

The objective function in the $\boldsymbol{\Theta}$-subproblem (12) has a quadratic form, whose solution is directly given by

$$\boldsymbol{\Theta}^{k+1} = \frac{\mathbf{S} + \rho \mathbf{G}^T \mathbf{V}^{k+1}}{2\rho + 1}. \quad (15)$$

Next, we give a result for the convergence of the ADM algorithm (10)-(12) for generalized penalty functions. While the convergence properties of ADM have been extensively studied for the convex case, there have been only a few studies for the nonconvex case. Very recently, the convergence of ADM has been analyzed under nonconvex frameworks in [29, 30]. The following result is derived via adopting the approaches in [29, 30].

**Theorem 1.** Suppose that $g_\lambda$ is a closed, proper, lower semi-continuous, KL function. Let $\mathbf{Z}^k = (\boldsymbol{\Theta}^k, \mathbf{V}_1^k, \mathbf{V}_2^k)$, for arbitrary starting point $\mathbf{Z}^0$, the sequence $\{\mathbf{Z}^k\}$ generated by the ADM algorithm via (10), (11) and (12) has finite length

$$\sum_{k=1}^\infty \|\mathbf{Z}^{k+1} - \mathbf{Z}^k\|_F < \infty. \quad (16)$$

In particular, the sequence $\{\mathbf{Z}^k\}$ converges to a stationary point of the problem (9).

*Proof*: See Appendix A.

In the proposed algorithm, a large value of the penalty parameter is desirable in order to enforce that $\|\mathbf{G}\boldsymbol{\Theta} - \mathbf{V}\|_F^2 \approx 0$. When $\rho \to \infty$, the solution of (9) accurately approaches that of the problem (8). However, an ADM algorithm in general tends to be very slow when $\rho$ gets very large. In practical applications, a standard trick to speed up the algorithm is to adopt a continuation process for the penalty parameter. Specifically, we can use a properly small starting value of the penalty parameter and gradually increase it by iteration until reaching the target value, e.g., $0 < \rho_0 \leq \rho_1 \leq \cdots \leq \rho_K = \rho_{K+1} = \cdots = \rho$. In this case, Theorem 1 still applies as the penalty parameter becomes fixed (at $\rho$) after a finite number of iterations. Moreover, in the case of a nonconvex penalty, the performance of the proposed algorithm is closely related to the initialization. Intensive numerical studies show that, the new algorithm can achieve satisfactory performance with an initialization by a convex $\ell_1$-penalty based method.

## IV. STATISTICAL PROPERTIES

In this section, we analyze the statistical properties of the proposed estimator for generalized nonconvex penalties. Similar to [20], we consider the following class of "approximately sparse" correlation matrices

$$\mathrm{M}(p, M_d, \xi) := \left\{\boldsymbol{\Theta} : \max_i \sum_{j \neq i} |\boldsymbol{\Theta}_{ij}|^p \leq M_d, \boldsymbol{\Theta}_{ii} = 1, \varphi_{\min}(\boldsymbol{\Theta}) = \xi\right\} \quad (17)$$

for $0 \leq p < 1$. Further, we define a class of covariance matrices based on (17) as

$$U(\kappa, p, M_d, \xi) := \left\{\boldsymbol{\Sigma} : \max_i \boldsymbol{\Sigma}_{ii} \leq \kappa, \tilde{\boldsymbol{\Sigma}}^{-1} \boldsymbol{\Sigma} \tilde{\boldsymbol{\Sigma}}^{-1} \in \mathrm{M}(p, M_d, \xi)\right\}$$

where $\tilde{\boldsymbol{\Sigma}} = \mathrm{diag}(\sqrt{\boldsymbol{\Sigma}_{11}}, \cdots, \sqrt{\boldsymbol{\Sigma}_{dd}})$.

To derive the statistical properties, we assume each marginal distribution of $\mathbf{X} := (X_1, \cdots, X_d)$ is sub-Gaussian and satisfies the exponential-tail condition

$$E\{\exp(tX_i^2)\} \leq K < \infty \quad \text{for} \quad |t| \leq \tau. \quad (18)$$

We first give a result for the special case of strictly sparse covariance matrices with $p = 0$, under the assumption that the penalty $g_\lambda$ satisfies

$$\max_{(i,j) \in S} g'_\lambda(\boldsymbol{\Theta}_{ij}^0) = O\left(\sqrt{\frac{\log d}{n}}\right) \quad \text{and} \quad \max_{(i,j) \in S} g''_\lambda(\boldsymbol{\Theta}_{ij}^0) = o(1) \quad (19)$$

where $\boldsymbol{\Theta}^0$ is the true correlation matrix, $S$ denotes the off-diagonal support of $\boldsymbol{\Theta}^0$, i.e., $S := \{(i,j) \in \mathbb{R}^{d \times d} : \boldsymbol{\Theta}_{ij}^0 \neq 0, i \neq j\}$. For example, if the penalty parameter $\lambda$ is selected to satisfy

$$\min_{(i,j) \in S} \frac{|\boldsymbol{\Theta}_{ij}^0|}{\lambda} \to \infty$$

as $n \to \infty$, for hard-thresholding and SCAD penalties, $\max_{(i,j) \in S} g'_\lambda(\boldsymbol{\Theta}_{ij}^0) = \max_{(i,j) \in S} g''_\lambda(\boldsymbol{\Theta}_{ij}^0) = 0$ for sufficiently large $n$, while for $\ell_1$-penalty $\max_{(i,j) \in S} g'_\lambda(\boldsymbol{\Theta}_{ij}^0) = \lambda$ and $\max_{(i,j) \in S} g''_\lambda(\boldsymbol{\Theta}_{ij}^0) = 0$.

**Theorem 2:** Suppose that $\boldsymbol{\Sigma}^0 \in U(\kappa, 0, M_d, \xi_{\min})$ for some $\xi_{\min} > 0$, and let $s = |S|$ denote the number of nonzero off-diagonal elements in $\boldsymbol{\Sigma}^0$. Under condition (18) and for a nonconvex penalty satisfies (19), if $\varepsilon_1 \leq \xi_{\min}$, $\lambda = O(\sqrt{s \log d / n})$ and $\sqrt{\log d / n} = o(1)$, then there at least exists a local minimizer $\hat{\boldsymbol{\Theta}}$ of (6) such that

$$\|\hat{\boldsymbol{\Theta}} - \boldsymbol{\Theta}^0\|_F = O_P\left(\sqrt{\frac{s \log d}{n}}\right).$$

Further, for the spectral norm of the covariance matrix $\hat{\boldsymbol{\Sigma}} = \mathrm{diag}(\mathbf{R})^{1/2} \hat{\boldsymbol{\Theta}} \mathrm{diag}(\mathbf{R})^{1/2}$,

$$\|\hat{\boldsymbol{\Sigma}} - \boldsymbol{\Sigma}^0\|_2 = O_P\left(\sqrt{\frac{s \log d}{n}}\right).$$

In addition, for the $\ell_1$-norm penalty, we only need $\lambda = O(\sqrt{\log d / n})$.

*proof:* See appendix B.

Note that this result allows $d \gg n$ as long as $\sqrt{\log d/n} = o(1)$, and it achieves the minimax optimal rate of convergence under both Frobenius and spectral norms [11]. This result is derived for the procedure that, the covariance matrix is obtained by estimating the correlation matrix firstly. If we estimate the covariance directly, the result rate of convergence would be $O_P(\sqrt{(d+s)\log d/n})$, where the part $\sqrt{d \log d/n}$ comes from estimating the diagonal.

Next, we give an asymptotic result for the generalized case of approximately sparse covariance matrices with $0 \leq p < 1$. Similar to [15], we assume that the thresholding\shrinkage operator $T_\lambda$ corresponding to the penalty $g_\lambda$ satisfies

(i) $T_\lambda(x) \leq \text{sign}(x) T_\lambda(|x|)$ and $|T_\lambda(x)| \leq |x|$;

(ii) $T_\lambda(x) = 0$ for $x \leq \lambda$;

(iii) $|T_\lambda(x) - x| \leq |\lambda|$. (20)

These conditions establish the sign consistency, shrinkage, thresholding, and limited shrinkage properties for the thresholding-shrinkage operator corresponding to a generalized penalty.

**Theorem 3:** Suppose that $\mathbf{\Sigma}^0 \in U(\kappa, p, M_d, \xi_{\min})$ for some $\xi_{\min} > 0$, and the thresholding\shrinkage operator corresponding to $g_\lambda$ satisfies (20). Let $\hat{\mathbf{\Theta}}$ denote the estimate given by (6), under condition (20), there exist constants $c_0$ and $c_1$ such that, if $\varepsilon_1 < \xi_{\min}$, $\lambda = c_0 \sqrt{\log d/n} = o(1)$ and $n \geq [(c_1 M_d)/(\xi_{\min} - \varepsilon_1)]^{2/(1-q)} \log d$, then

$$\|\hat{\mathbf{\Theta}} - \mathbf{\Theta}^0\|_F = O_P\left(M_d \left(\frac{\log d}{n}\right)^{\frac{1-q}{2}}\right).$$

Further, for the spectral norm of the covariance matrix $\hat{\mathbf{\Sigma}} = \text{diag}(\mathbf{R})^{1/2} \hat{\mathbf{\Theta}} \text{diag}(\mathbf{R})^{1/2}$,

$$\|\hat{\mathbf{\Sigma}} - \mathbf{\Sigma}^0\|_2 = O_P\left(M_d \left(\frac{\log d}{n}\right)^{\frac{1-q}{2}}\right).$$

Theorem 3 is derived based on the results in [15]. Specifically, under the conditions in Theorem 3, it can be shown that the estimator (6) gives the same estimation as the generalized thresholding estimator (1) with overwhelming probability in the asymptotic case. The detailed proof follows similarly to that in [20] with some minor changes, which is omitted here for succinctness. Note that, in the strictly sparse case of $q = 0$, $M_d$ is a bound on the number of non-zero elements in each row (without counting the diagonal element). Thus, in this case $\sqrt{s} \approx M_d$, and the rate in Theorem 3 coincides with that in Theorem 2, even though the approaches of proof are very different.

In finite sample cases, the generalized thresholding estimator (1) is not guaranteed to be positive-definite, but, empirically, it can be positive-definite in many situations. Thus, in practical applications, we can first check the minimum eigenvalue of the generalized thresholding estimator, and proceed to the proposed algorithm only when the feasibility condition of positive-definiteness is not satisfied.

## V. EXTENSION TO COVARIANCE SKETCHING

Sketching via randomly linear measurements is a useful algorithmic tool for dimensionality reducing, which has been widely used in computer science [32, 33] and compressed sensing [34, 35]. Recently, covariance estimation from *sketches* or *compressed measurements* has attracted much attention in various fields of science and engineering [36-42]. An important application of covariance sketching arises in Big Data, e.g., covariance estimation from high-dimensional data stream. To extract the covariance information from high-dimensional real-time data stream at a high rate, it may be infeasible or undesirable to sample and store the whole stream due to memory and processing power constraint. In such a scenario, it is desirable to extract the covariance information of the data instances from compressed inputs with low requirement of memory and computational complexity and without storing the entire stream.

Let $\mathbf{x} \in \mathbb{R}^d$ denote a zero-mean random vector with covariance matrix $\mathbf{\Sigma}^0$, $\mathbf{A} \in \mathbb{R}^{m \times d}$ is the sampling matrix with $m < d$, with compressed samples $\mathbf{y}_t = \mathbf{A} \mathbf{x}_t$, $t = 1, 2, \cdots, n$, define the sample covariance of $\mathbf{y}_t$ as

$$\begin{aligned} \mathbf{Y} &= \frac{1}{n} \sum_{t=1}^{n} \mathbf{y}_t^T \mathbf{y}_t = \mathbf{A} \mathbf{R} \mathbf{A}^T \\ &= \mathbf{A} \mathbf{\Sigma}^0 \mathbf{A}^T + \underbrace{\mathbf{A}(\mathbf{R} - \mathbf{\Sigma}^0) \mathbf{A}^T}_{\mathbf{N}}. \end{aligned} \quad (21)$$

Under the assumption that $\mathbf{x}$ is zero-mean and its covariance is finite, the perturbation $\mathbf{N}$ in (21) satisfies $E\{\mathbf{N}\} = \mathbf{0}$ and $E\{\|\mathbf{N}\|_F^2\} \propto 1/n$. Since $m < d$, the reconstruction of $\mathbf{\Sigma}^0$ from $\mathbf{Y}$ is an ill-posed problem. However, if the covariance matrix $\mathbf{\Sigma}^0$ has low-dimensional structures such as low-rankness and sparsity, it can be accurately reconstructed from $\mathbf{Y}$ by exploiting such structural information. Given the sample covariance $\mathbf{Y}$ of the compressed observations, we can find a positive definite estimation of $\mathbf{\Sigma}^0$ via solving the following optimization problem

$$\min_{\mathbf{\Sigma}} \frac{1}{2} \|\mathbf{Y} - \mathbf{A} \mathbf{\Sigma} \mathbf{A}^T\|_F^2 + g_\lambda(\mathbf{\Sigma}) \quad \text{subject to} \quad \mathbf{\Sigma} \succeq \varepsilon_2 \mathbf{I}_d \quad (22)$$

where $\varepsilon_2 > 0$ is the lower bound for the minimum eigenvalue of the covariance matrix. When $\mathbf{A} = \mathbf{I}_d$ and the covariance matrix is replaced by the correlations matrix, (22) reduces to the correlation estimation problem (5). Very recently, an ALM algorithm has been proposed in [42] to solve (22) for convex penalties, e.g., $g_\lambda(\Sigma)=\|\Sigma\|_1$. However, this algorithm is not guaranteed to converge for a nonconvex $g_\lambda$. In this following, we show that the ADM algorithm presented in section III can be extended to solve (22) with convergence guarantee for a nonconvex penalty.

Similar to (7), the problem (22) can be equivalently rewritten as

$$\min_\Sigma \frac{1}{2}\|\mathbf{Y}-\mathbf{A}\Sigma\mathbf{A}^T\|_F^2 + g_\lambda(\Sigma) + \delta_{\Omega_2}(\Sigma). \quad (23)$$

where $\Omega_2:=\{\Sigma:\Sigma\geq\varepsilon_2\mathbf{I}_d\}$ is a closed convex set, $\delta_{\Omega_2}(\cdot)$ denotes the indicator function defined on this set. Using two auxiliary variables $\Gamma_1,\Gamma_2\in\mathbb{R}^{d\times d}$, the problem (23) can be equivalently rewritten as

$$\min_{\Sigma,\Gamma_1,\Gamma_2} \frac{1}{2}\|\mathbf{Y}-\mathbf{A}\Sigma\mathbf{A}^T\|_F^2 + g_\lambda(\Gamma_1) + \delta_{\Omega_2}(\Gamma_2)$$
$$\text{subject to} \quad \mathbf{G}\Sigma=\Gamma \quad (24)$$

with

$$\Gamma = \begin{bmatrix} \Gamma_1 \\ \Gamma_2 \end{bmatrix}.$$

The constrained minimization problem (24) can be attacked via solving

$$\min_{\Sigma,\Gamma_1,\Gamma_2} \frac{1}{2}\|\mathbf{Y}-\mathbf{A}\Sigma\mathbf{A}^T\|_F^2 + g_\lambda(\Gamma_1) + \delta_{\Omega_2}(\Gamma_2) + \frac{\rho}{2}\|\mathbf{G}\Sigma-\Gamma\|_F^2 \quad (25)$$

where $\rho>0$ is a penalty parameter. For sufficiently large $\rho$, e.g., $\rho\to+\infty$, the solution of (25) accurately approaches that of the problem (24). Then, ADM applied to (25) consists of the following three steps in the $k+1$-th iteration

$$\Gamma_1^{k+1} = \arg\min_{\Gamma_1}\left(g_\lambda(\Gamma_1) + \frac{\rho}{2}\|\Sigma^k-\Gamma_1\|_F^2 + \frac{c_k}{2}\|\Gamma_1-\Gamma_1^k\|_F^2\right) \quad (26)$$

$$\Gamma_2^{k+1} = \arg\min_{\Gamma_2}\left(\delta_{\Omega_2}(\Gamma_2) + \frac{\rho}{2}\|\Sigma^k-\Gamma_2\|_F^2 + \frac{d_k}{2}\|\Gamma_2-\Gamma_2^k\|_F^2\right) \quad (27)$$

$$\Sigma^{k+1} = \arg\min_\Sigma\left(\frac{1}{2}\|\mathbf{Y}-\mathbf{A}\Sigma\mathbf{A}^T\|_F^2 + \frac{\rho}{2}\|\mathbf{G}\Sigma-\Gamma^{k+1}\|_F^2\right). \quad (28)$$

The $\Gamma_1$- and $\Gamma_2$-subproblems can be respectively solved as

$$(\Gamma_1^{k+1})_{ij} = \begin{cases} T_{\lambda/(\rho+c_k)}\left(\dfrac{\rho\Sigma_{ij}^k + c_k(\Gamma_1^k)_{ij}}{\rho+c_k}\right), & i\neq j \\ \Sigma_{ii}^k, & i=j \end{cases} \quad (29)$$

and

$$\Gamma_2^{k+1} = \mathrm{P}_+\left(\frac{\rho\Sigma^k + d_k\Gamma_2^k}{\rho+d_k}, \varepsilon_2\right). \quad (30)$$

Take the orthogonal eigen-decomposition of $\mathbf{A}^T\mathbf{A}$ as $\mathbf{A}^T\mathbf{A} = \mathbf{E}\Lambda\mathbf{E}^T$, the solution of the $\Sigma$-subproblem (28) is given by (see Appendix C)

$$\Sigma^{k+1} = \mathbf{E}[(\mathbf{E}^T(\mathbf{A}^T\mathbf{Y}\mathbf{A}+\rho\mathbf{G}^T\Gamma^{k+1})\mathbf{E})\oslash(\mathbf{a}\mathbf{a}^T+2\rho\mathbf{1}_{n\times n})]\mathbf{E}^T \quad (31)$$

where the vector $\mathbf{a}\in\mathbb{R}^d$ contains the eigenvalues of $\mathbf{A}^T\mathbf{A}$, i.e., $\Lambda=\mathrm{diag}(\mathbf{a})$.

**Theorem 4.** Suppose that $g_\lambda$ is a closed, proper, lower semi-continuous, KL function. Let $\mathbf{Z}^k=(\Sigma^k,\Gamma_1^k,\Gamma_2^k)$, for arbitrary starting point $\mathbf{Z}^0$, the sequence $\{\mathbf{Z}^k\}$ generated by the ADM algorithm via (26), (27) and (28) has finite length

$$\sum_{k=1}^\infty \|\mathbf{Z}^{k+1}-\mathbf{Z}^k\|_F < \infty. \quad (32)$$

In particular, the sequence $\{\mathbf{Z}^k\}$ converges to a stationary point of the problem (25).

*Proof*: See Appendix A.

In implementing the proposed algorithm, a large value of $\rho$ is desirable in order to enforce that $\|\mathbf{G}\Sigma-\Gamma\|_F^2 \approx 0$. Similar to the discussion in section III, we can adopt a continuation process for the penalty parameter to speed up this algorithm. Moreover, for a nonconvex penalty, a good initialization, e.g., by a convex $\ell_1$-penalty based method, is crucial for this algorithm to achieve satisfactory performance.

## VI. NUMERICAL EXPERIMENTS

We evaluate the proposed estimators using the SCAD, hard-, $\ell_q$-, and lasso penalties, termed as SCAD-ADM, Hard-ADM, Lq-ADM and L1-ADM, respectively, in comparison with the ALM algorithm using the lasso penalty [20], termed L1-ALM. We chose $q=0.5$ for Lq-ADM and set a lower bound of $10^{-3}$ for the minimum eigenvalue for each estimator. Moreover, for the proposed algorithms, a continuation process is used for the penalty parameter as $\rho_k=1.09\rho_{k-1}$ if $\rho_k<\rho$ and $\rho_k=\rho$ otherwise. The thresholding parameter for each estimator is chosen by subsampling and fivefold cross-validation [13]. We conduct mainly two evaluation experiments on simulated data sets and a real gene data set, respectively.

Matlab codes for the proposed estimators and for reproducing the results in the following experiments are available at **https://github.com/FWen/Nonconvex-PDLCE.git**.

### A. Simulated Datasets

We consider three typical sparse covariance matrix models, which are standard test cases in the literature. Fig. 2 shows the heat maps of these there covariance models with $d=100$.

*Block matrix:* Partition the indices $\{1,2,\cdots,d\}$ evenly into $K=d/20$ nonoverlapping subsets $S_1,\cdots,S_K$, the covariance is given by

$$(\Sigma^0)_{ij} = 0.2 I(i=j) + 0.8\sum_{k=1}^K I(i\in S_k, j\in S_k).$$

*Toeplitz matrix:*

$$(\mathbf{\Sigma}^0)_{ij} = 0.75^{|i-j|}.$$

*Banded matrix:*

$$(\mathbf{\Sigma}^0)_{ij} = \left(1 - |i-j|/10\right)_+.$$

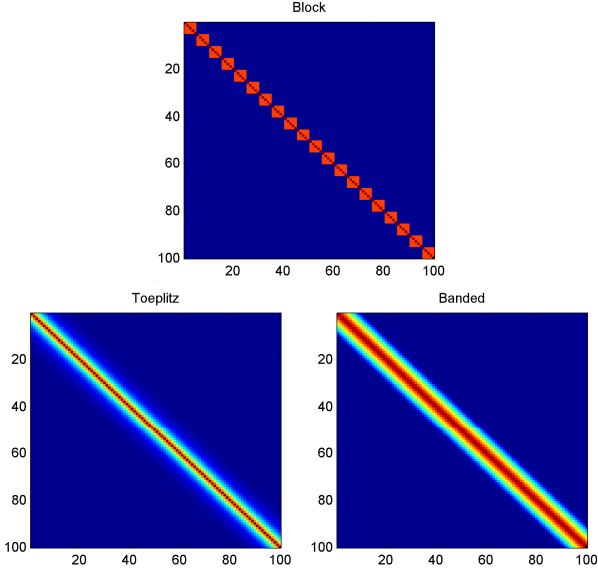

Fig. 2. Heat maps of three simulated covariance matrices for $d=100$.

We consider two dimension conditions with $d=100$ and $d=400$, respectively, and five sample number conditions of $n \in \{100, 200, 400, 600, 800\}$. The performance is evaluated in terms of the relative error of estimation under both the Frobenius norm and the spectral norm. Each provided result is an average over 100 independent Monte Carlo runs. For each independent run, the data set (with a size of $n$) is generated from a $d$-dimensional Gaussian random variable with zero-mean and covariance $\mathbf{\Sigma}^0$.

Fig. 3 and Fig. 4 show the estimation performance of the compared estimators for $d=100$ and $d=400$, respectively. In the condition of $d=100$, the proposed estimators with the nonconvex SCAD, hard- and $\ell_q$-penalties show considerable performance gain over the lasso penalty based L1-ALM and L1-ADM estimators, except for the case of Toeplitz covariance model under the Frobenius norm metric. This is due to the fact that the considered Toeplitz covariance model is less sparse than the other two covariance models. For example, for the block covariance model and $n=200$, the averaged estimation errors of the SCAD-ADM, Hard-ADM and Lq-ADM estimators under the Frobenius norm are approximately 57.3%, 59.7% and 62.3% that of the L1-ALM estimator, while that under the spectral norm are 66.4%, 69.5% and 74.2%, respectively. The advantages of SCAD-ADM, Hard-ADM and Lq-ADM estimators over the L1-ALM and L1-ADM estima-tors are more significant in the condition of $d=400$, since the three covariance models in this condition are more sparse compared with the condition of $d=100$. For example, in this condition the block covariance model and $n=200$, the averaged estimation errors of the SCAD-ADM, Hard-ADM and Lq-ADM estimators under the Frobenius norm are approximately 59.5%, 59.1% and 61.3% that of the L1-ALM estimator, while that under the spectral norm are 66.3%, 67.5% and 67.3%, respectively. The proposed estimator with lasso penalty (L1-ADM) performs comparably with the L1-ALM estimator. Fig. 5 shows the eigenvalues of the compared estimators in a typical case for $d=100$ and $n=200$. Extensive numerical studies show that all the compared estimators can achieve a positive definiteness rate of 100%, but the generalized thresholding estimators often yield indefinite covariance matrices, similar to the results shown in [18-20].

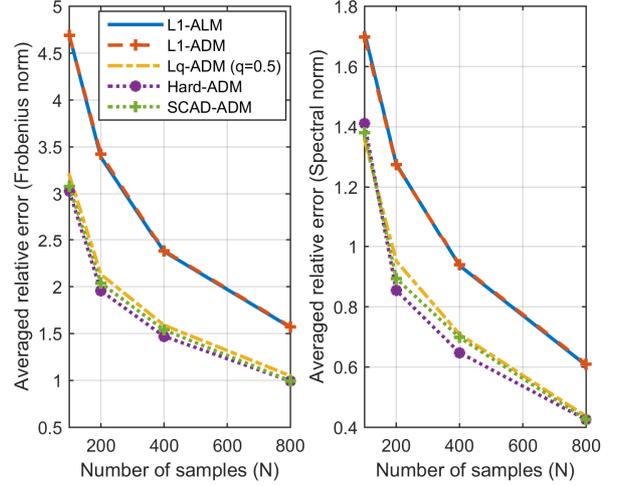

(a) Block matrix

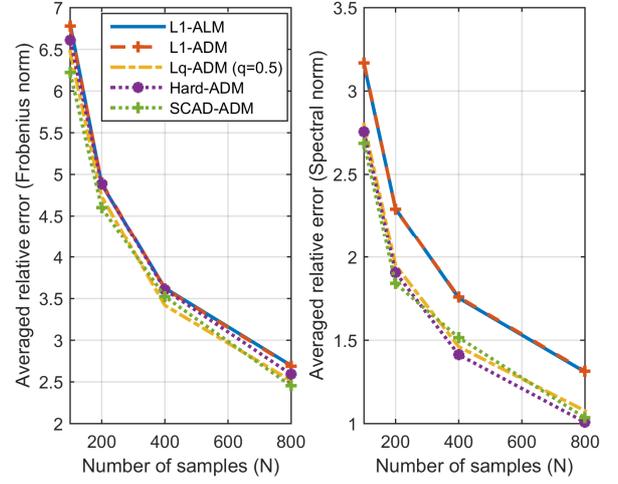

(b) Toeplitz matrix

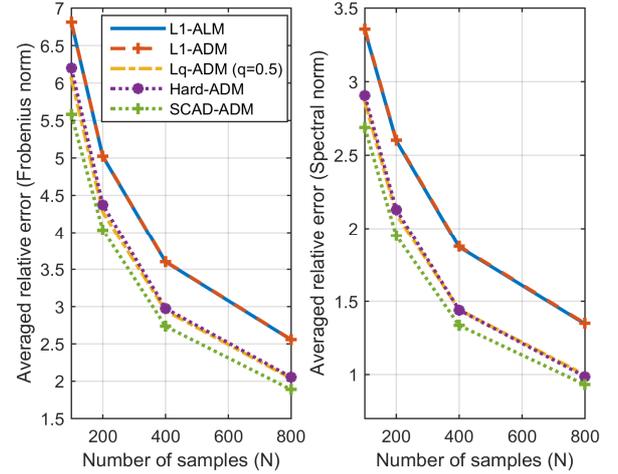

(c) Banded matrix

Fig. 3. Estimation performance of the compared algorithms for $d=100$.

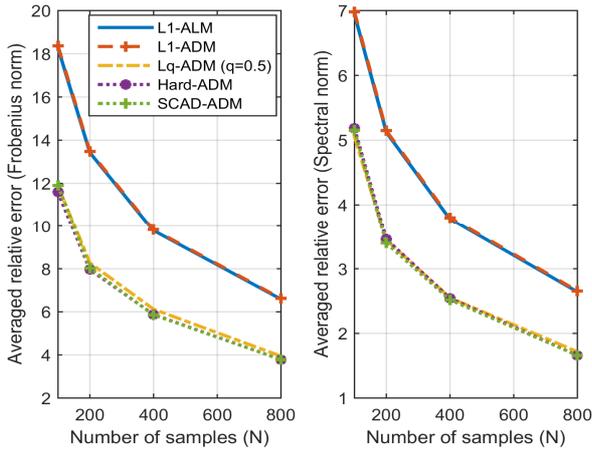

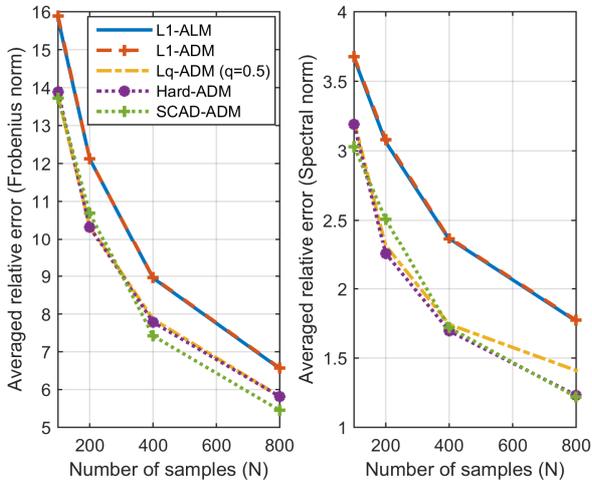

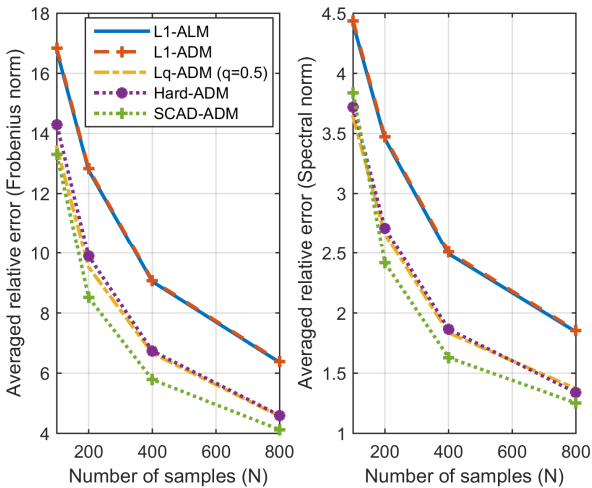

(a) Block matrix

(b) Toeplitz matrix

(c) Banded matrix

Fig. 4. Estimation performance of the compared algorithms for $d=400$.

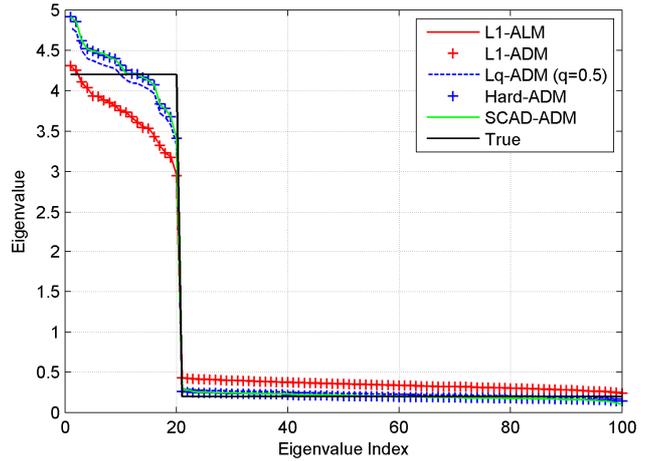

Fig. 5. Typical plots of the eigenvalues of the compared estimators for $d=100$ and $n=200$.

### B. Gene Clustering Example

Gene clustering based on the correlations among genes is a popular technique in gene expression data analysis [43]. Here we consider a gene clustering example using a gene expression dataset from a small round blue-cell tumors (SRBCTs) microarray experiment [44] to further evaluate the compared methods. This dataset contains 88 SRBCT tissue samples, with 63 labeled calibration samples and 25 test samples, and 2308 gene expression values are recorded for each of the samples. These 2308 genes were selected out from 6567 originally measured genes via requiring that each gene has a red intensity greater than 20 over all the tissue samples. We use the 63 labeled calibration samples and pick up the top 40 and bottom 160 genes based on their F-statistic as done in [15]. Accordingly, the top 40 genes are informative while the bottom 160 genes are non-informative, and the dependence between these two parts is weak. We apply hierarchical clustering (group average agglomerative clustering) to the genes using the estimated correlations by the compared methods. The generalized thresholding estimators [15], including soft-, hard- and SCAD-thresholding, are also tested in the comparison. Fig. 6 shows the heat map of the gene expression data of the top 40 genes of the 63 samples. The genes are sorted by hierarchical clustering based on the sample correlations, and the samples are sorted by tissue class (there are four classes of tumors in the sample, including 23 EWS, 8 BL, 12 NB, and 20 RMS).

Fig. 7 shows the heat maps of the absolute values of estimated correlations by the compared estimators for the selected 200 genes. For each estimator, the presented heat map is ordered by group average agglomerative clustering based on the estimated correlation matrix. Meanwhile, for each estimated correlation matrix, the percentage of the entries with absolute values less than $10^{-5}$ is also shown. Fig. 8 plots the bottom 80 eigenvalues (ordered in descending values) of the compared estimators. It can be clearly seen from the heat maps that, compared with the L1-ALM and L1-ADM estimators, the SCAD-ADM, Hard-ADM and Lq-ADM estimators give cleaner and more informative estimates of the sparsity pattern. Moreover, the soft-, hard- and SCAD-thresholding estimators yield cleaner estimates of the sparsity pattern than the iterative positive estimators. However, from Fig. 8, while the estimates of the proposed methods and the L1-ALM method are positive-definite, the estimates of soft-, hard- and SCAD-thresholding methods are indefinite. Specifically, these three generalized thresholding estimators contain 13, 40 and 10 negative eigenvalues, respectively.

## VII. Conclusion

This work proposed a class of positive-definite covariance estimators using generalized nonconvex penalties, and developed an alternating direction algorithm to efficiently solve the corresponding formulation. The proposed algorithm is guaranteed to converge in the case of a generalized nonconvex penalty. The established statistical properties of the new estimators indicate that, for generalized nonconvex penalties, they can attain the optimal rates of convergence in terms of both the Frobenius norm and spectral norm errors. Extension of the proposed algorithm to covariance sketching has been discussed. In the simulation study, the new estimators showed significantly lower estimation error under both the Frobenius norm and the spectral norm compared with lasso penalty based estimators. The advantage of the new estimators has also been demonstrated by a gene clustering example using a gene expression dataset from a small round blue-cell tumors microarray experiment.

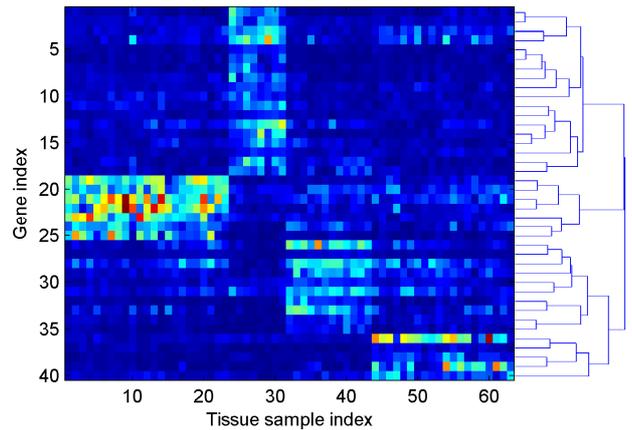

(a) Gene expression data

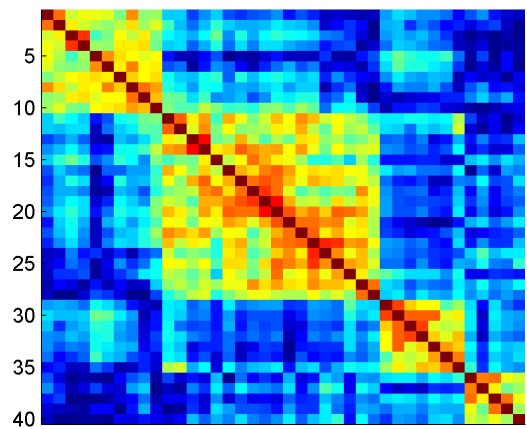

(b) Sample correlations

Fig. 6. Heat maps of the gene expression data and sample correlations of the top 40 genes (with genes sorted by hierarchical clustering and tissue samples sorted by tissue class).

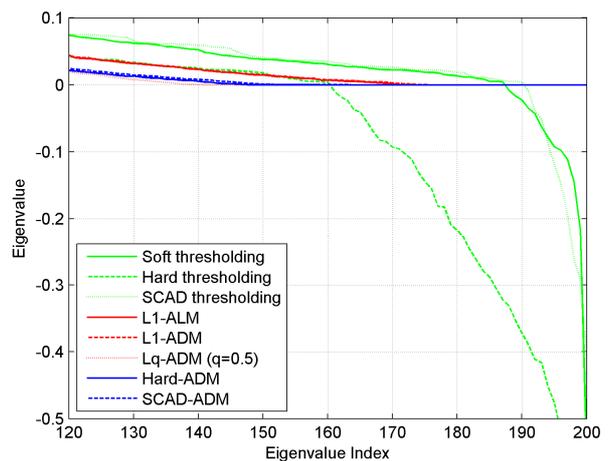

Fig. 8. Plots of the bottom 80 eigenvalues of the compared estimators.

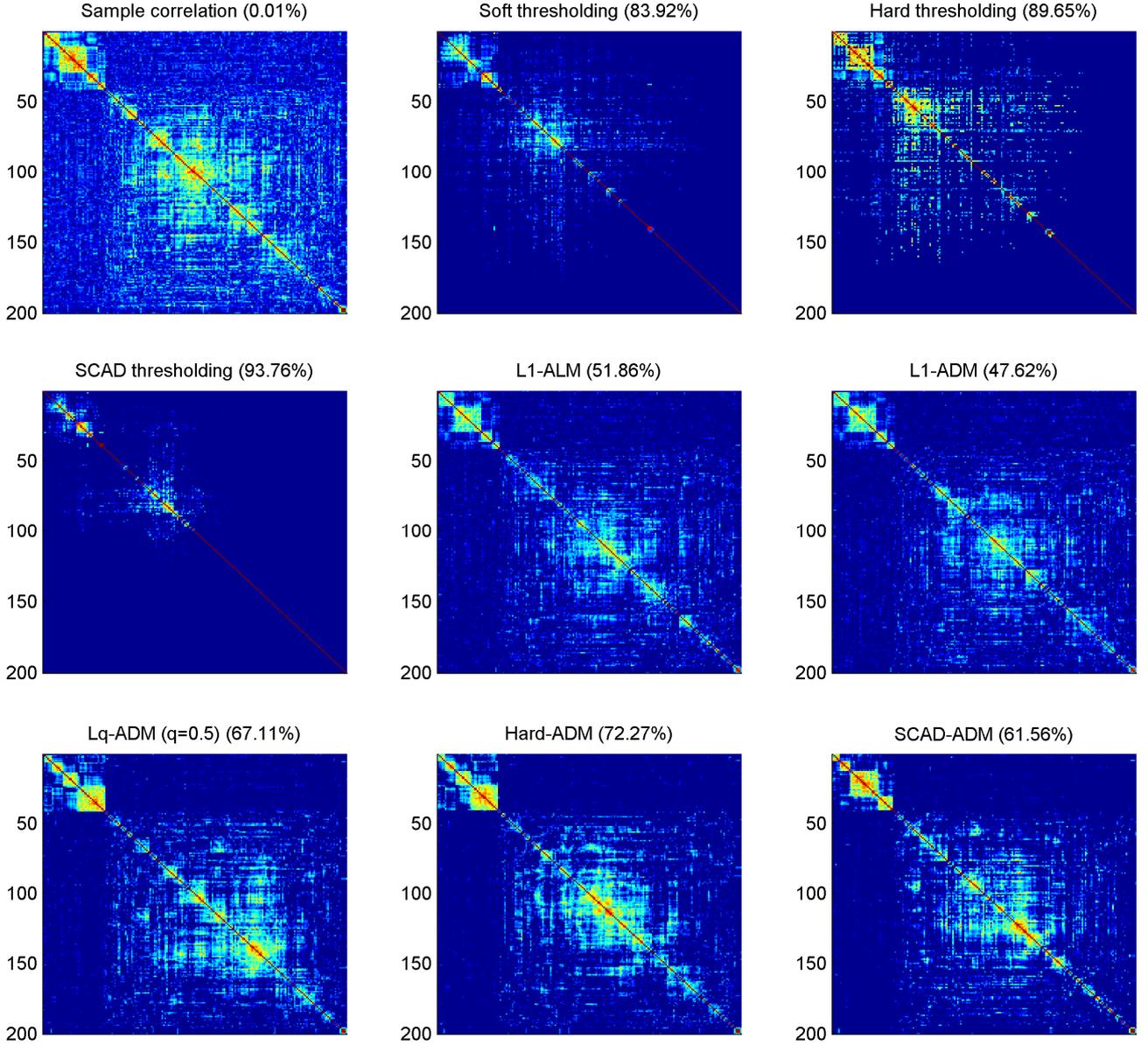

Fig. 7. Heat maps of the absolute values of estimated correlations by the compared estimators for the selected 200 genes. The percentage of the entries with absolute values less than $10^{-5}$ is given in parentheses.

## APPENDIX A

Note that, the ADM algorithm (10)-(12) is a special case of the algorithm (26)-(28) when $\mathbf{A} = \mathbf{I}_d$ and the covariance matrix is replaced by the correlations matrix. Accordingly, we only present the proof of Theorem 4, while Theorem 1 can be derived in a similar manner. Theorem 4 is derived via adopting the approach in [29, 30], we only sketch the proof here. In the sequel for convenience we use the notations

$$L(\mathbf{\Sigma}, \mathbf{\Gamma}_1, \mathbf{\Gamma}_2) := \frac{1}{2}\|\mathbf{Y} - \mathbf{A}\mathbf{\Sigma}\mathbf{A}^T\|_F^2 \\ + g_\lambda(\mathbf{\Gamma}_1) + \delta_{\Omega_2}(\mathbf{\Gamma}_2) + \frac{\rho}{2}\|\mathbf{G}\mathbf{\Sigma} - \mathbf{\Gamma}\|_F^2 \quad (33)$$

and

$$\mathbf{Z}^k := (\mathbf{\Sigma}^k, \mathbf{\Gamma}_1^k, \mathbf{\Gamma}_2^k).$$

We first give the following lemmas in the proof of Theorem 4.

**Lemma 1.** Let $\{\mathbf{Z}^k\}$ be a sequence generated by the ADM algorithm (26)-(28), then the sequence $\{L(\mathbf{Z}^k)\}$ is nonincreasing and

$$\sum_{k=0}^{\infty} \|\mathbf{Z}^{k+1} - \mathbf{Z}^k\|_F^2 < \infty$$

and hence $\lim_{k \to \infty} \|\mathbf{Z}^{k+1} - \mathbf{Z}^k\|_F = 0$.

**Lemma 2.** Let $\{\mathbf{Z}^k\}$ be a sequence generated by the ADM algorithm (26)-(28), define

$$\mathbf{B}_{\Gamma_1}^k = c_{k-1}(\Gamma_1^{k-1} - \Gamma_1^k) - \rho(\Sigma^{k-1} - \Sigma^k)$$

$$\mathbf{B}_{\Gamma_2}^k = d_{k-1}(\Gamma_2^{k-1} - \Gamma_2^k) - \rho(\Sigma^{k-1} - \Sigma^k).$$

Then, $(\mathbf{0}, \mathbf{B}_{\Gamma_1}^k, \mathbf{B}_{\Gamma_2}^k) \in \partial L(\mathbf{Z}^k)$ and there exists $C_1 > 0$ such that

$$\|(\mathbf{0}, \mathbf{B}_{\Gamma_1}^k, \mathbf{B}_{\Gamma_2}^k)\|_F \leq C_1 \|\mathbf{Z}^k - \mathbf{Z}^{k-1}\|_F.$$

**Lemma 3.** Let $\{\mathbf{Z}^k\}$ be a sequence generated by the ADM algorithm (26)-(28), and denote the cluster point set by $\omega$. The following assertions hold.

(i) Any cluster point of $\{\mathbf{Z}^k\}$ is a stationary point of $L$.

(ii) $\lim_{k \to \infty} \text{dist}(\mathbf{Z}^k, \omega) = 0$ and $\omega$ is a nonempty, compact and connected set.

(iii) The objective function $L$ is finite and constant on $\omega$.

**Proof of Lemma 1:** From the definition of $\Gamma_1^{k+1}$ as a minimizer of the objective in (26), we have

$$L(\Sigma^k, \Gamma_1^{k+1}, \Gamma_2^k) + \frac{c_k}{2} \|\Gamma_1^{k+1} - \Gamma_1^k\|_F^2 \leq L(\Sigma^k, \Gamma_1^k, \Gamma_2^k). \quad (34)$$

Similarly, it follows from (27) that

$$L(\Sigma^k, \Gamma_1^{k+1}, \Gamma_2^{k+1}) + \frac{d_k}{2} \|\Gamma_2^{k+1} - \Gamma_2^k\|_F^2 \leq L(\Sigma^k, \Gamma_1^{k+1}, \Gamma_2^k). \quad (35)$$

Moreover, the Hessian of $L(\Sigma, \Gamma_1^{k+1}, \Gamma_2^{k+1})$ satisfies

$$\nabla^2 L(\Sigma, \Gamma_1^{k+1}, \Gamma_2^{k+1}) = (\mathbf{A}^T \mathbf{A}) \otimes (\mathbf{A}^T \mathbf{A}) + 2\rho \mathbf{I}_{d^2} \geq [\varphi_{\min}^2(\mathbf{A}^T \mathbf{A}) + 2\rho] \mathbf{I}_{d^2} \quad (36)$$

where the inequality follows from $(\mathbf{A}^T\mathbf{A}) \otimes (\mathbf{A}^T\mathbf{A}) \leq \varphi_{\min}^2(\mathbf{A}^T\mathbf{A})\mathbf{I}_{d^2}$ since the eigenvalues of $\mathbf{A} \otimes \mathbf{B}$ are the pairwise products of the eigenvalues of $\mathbf{A}$ and $\mathbf{B}$. The inequality in (36) implies the objective function in the $\Sigma$-subproblem (28) is $(\varphi_{\min}^2(\mathbf{A}^T\mathbf{A}) + 2\rho)$-strongly convex with respect to $\Sigma$. Then, from the definition of $\Sigma^{k+1}$ as a minimizer, i.e., $\nabla L(\Sigma, \Gamma_1^{k+1}, \Gamma_2^{k+1}) = 0$, we have

$$L(\Sigma^{k+1}, \Gamma_1^{k+1}, \Gamma_2^{k+1}) + \frac{\varphi_{\min}^2(\mathbf{A}^T\mathbf{A}) + 2\rho}{2} \|\Sigma^{k+1} - \Sigma^k\|_F^2 \leq L(\Sigma^k, \Gamma_1^{k+1}, \Gamma_2^{k+1}) \quad (37)$$

Summing (34), (35) and (37), we obtain

$$L(\Sigma^k, \Gamma_1^k, \Gamma_2^k) - L(\Sigma^{k+1}, \Gamma_1^{k+1}, \Gamma_2^{k+1}) \geq \frac{c_k}{2} \|\Gamma_1^{k+1} - \Gamma_1^k\|_F^2 + \frac{d_k}{2} \|\Gamma_2^{k+1} - \Gamma_2^k\|_F^2 + \frac{\varphi_{\min}^2(\mathbf{A}^T\mathbf{A}) + 2\rho}{2} \|\Sigma^{k+1} - \Sigma^k\|_F^2 \quad (38)$$

which implies that $L(\Sigma^k, \Gamma_1^k, \Gamma_2^k)$ is nonincreasing. Follows from (38), we can find a positive constant $C_0$ such that

$$L(\Sigma^k, \Gamma_1^k, \Gamma_2^k) - L(\Sigma^{k+1}, \Gamma_1^{k+1}, \Gamma_2^{k+1}) \geq \frac{C_0}{2} \|\mathbf{Z}^{k+1} - \mathbf{Z}^k\|_F^2. \quad (39)$$

Let $N$ be a positive integer, summing up (39) from $k=0$ to $N-1$ we have

$$\sum_{k=0}^{N-1} \|\mathbf{Z}^{k+1} - \mathbf{Z}^k\|_F^2 \leq \frac{2}{C_0} (L(\mathbf{Z}^0) - L(\mathbf{Z}^N)).$$

Since $L$ is lower semi-continuous, it is bounded from below. Further, since $L(\mathbf{Z}^k)$ is nonincreasing, it converges to some real number $\overline{L}_\rho$. Taking the limit as $N \to \infty$, we obtain

$$\sum_{k=0}^{\infty} \|\mathbf{Z}^{k+1} - \mathbf{Z}^k\|_F^2 \leq \frac{2}{C_0} (L(\mathbf{Z}^0) - \overline{L}) < \infty.$$

**Proof of Lemma 2:** From the definition of the iterative steps (26)-(28), $(\Sigma^k, \Gamma_1^k, \Gamma_2^k)$ given by the $k$-th iteration satisfies

$$c_{k-1}(\Gamma_1^{k-1} - \Gamma_1^k) - \rho(\Sigma^{k-1} - \Gamma_1^k) \in \partial g_\lambda(\Gamma_1^k) \quad (40)$$

$$d_{k-1}(\Gamma_2^{k-1} - \Gamma_2^k) - \rho(\Sigma^{k-1} - \Gamma_2^k) \in \partial \delta_{\Omega_2}(\Gamma_2^k) \quad (41)$$

and

$$\mathbf{0} = \nabla_\Sigma L(\mathbf{Z}^k). \quad (42)$$

Then, from (40) and (41), it is easy to see

$$\mathbf{B}_{\Gamma_1}^k \in \partial_{\Gamma_1} L(\mathbf{Z}^k)$$

$$\mathbf{B}_{\Gamma_2}^k \in \partial_{\Gamma_2} L(\mathbf{Z}^k)$$

which together with (42) implies $(\mathbf{0}, \mathbf{B}_{\Gamma_1}^k, \mathbf{B}_{\Gamma_2}^k) \in \partial L(\mathbf{Z}^k)$. Furthermore, it is easy to see from (33) that the sequence $\{\mathbf{Z}^k\}$ is bounded, since all the sub-functions of $L$ are coercive. Thus, we have

$$\begin{aligned}
&\|(\mathbf{0}, \mathbf{B}_{\Gamma_1}^k, \mathbf{B}_{\Gamma_2}^k)\|_F \\
&\leq \|\mathbf{B}_{\Gamma_1}^k\|_F + \|\mathbf{B}_{\Gamma_2}^k\|_F \\
&\leq c_{k-1} \|\Gamma_1^{k-1} - \Gamma_1^k\|_F + d_{k-1} \|\Gamma_2^{k-1} - \Gamma_2^k\|_F + 2\rho \|\Sigma^{k-1} - \Sigma^k\|_F \\
&\leq C_1 \|\mathbf{Z}^k - \mathbf{Z}^{k-1}\|_F
\end{aligned} \quad (43)$$

for some $C_1 > 0$.

**Proof of Lemma 3:** Since $\{\mathbf{Z}^k\}$ is bounded, for a cluster point $\mathbf{Z}^* := (\Sigma^*, \Gamma_1^*, \Gamma_2^*)$, there exists a subsequence $\{\mathbf{Z}^{k_j}\}$ which converges to $\mathbf{Z}^*$. Since $g_\lambda(\Gamma_1)$ is lower semi-continuous, it follows that

$$\liminf_{j \to \infty} g_\lambda(\Gamma_1^{k_j}) \geq g_\lambda(\Gamma_1^*). \quad (44)$$

From the definition of $\mathbf{\Gamma}_1^{k_j}$ as a minimizer, it follows that $L(\mathbf{\Sigma}^{k_j-1}, \mathbf{\Gamma}_1^{k_j}, \mathbf{\Gamma}_2^{k_j-1}) \leq L(\mathbf{\Sigma}^{k_j-1}, \mathbf{\Gamma}_1^*, \mathbf{\Gamma}_2^{k_j-1})$. Then, taking limit as $j \to \infty$ we obtain

$$\limsup_{j \to \infty} g_\lambda(\mathbf{\Gamma}_1^{k_j}) + \frac{\rho}{2} \|\mathbf{\Sigma}^* - \mathbf{\Gamma}_1^*\|_F^2 \leq g_\lambda(\mathbf{\Gamma}_1^*) + \frac{\rho}{2} \|\mathbf{\Sigma}^* - \mathbf{\Gamma}_1^*\|_F^2$$

which together with (44) yields $\lim_{j \to \infty} g_\lambda(\mathbf{\Gamma}_1^{k_j}) = g_\lambda(\mathbf{\Gamma}_1^*)$. In a similar manner, we have $\lim_{j \to \infty} \delta_{\Omega_2}(\mathbf{\Gamma}_2^{k_j}) \geq \delta_{\Omega_2}(\mathbf{\Gamma}_2^*)$ and finally get

$$\lim_{j \to \infty} L(\mathbf{Z}^{k_j}) = L(\mathbf{Z}^*).$$

From Lemma 2, $(\mathbf{0}, \mathbf{B}_{\Gamma_1}^k, \mathbf{B}_{\Gamma_2}^k) \in \partial L(\mathbf{Z}^k)$ and $(\mathbf{0}, \mathbf{B}_{\Gamma_1}^k, \mathbf{B}_{\Gamma_2}^k) \to (\mathbf{0}, \mathbf{0}, \mathbf{0})$ as $k \to \infty$, which together with the closedness property of $\partial L$ imply that $(\mathbf{0}, \mathbf{0}, \mathbf{0}) \in \partial L(\mathbf{Z}^*)$. Thus, $\mathbf{Z}^*$ is a stationary point of $L$.

Properties (ii) is generic for any sequence $\{\mathbf{Z}^k\}$ which satisfies $\lim_{k \to \infty} \|\mathbf{Z}^{k+1} - \mathbf{Z}^k\|_F = \mathbf{0}$. The proof is given in Lemma 5 in [30]. Properties (iii) is straightforward since $L(\mathbf{Z}^k)$ is convergent (see Lemma 1).

**Proof of Theorem 4:** Based on the above lemmas, the rest proof of Theorem 4 is to show that the generated sequence $\{\mathbf{Z}^k\}$ has finite length, i.e.,

$$\sum_{k=0}^{\infty} \|\mathbf{Z}^{k+1} - \mathbf{Z}^k\|_F < \infty \tag{45}$$

which implies that $\{\mathbf{Z}^k\}$ is a *Cauchy sequence* and thus is convergent. Finally, this property together with Lemma 3 implies that $\{\mathbf{Z}^k\}$ converges to a stationary point of $L$. The derivation of (45) relies heavily on the KL property of $L$, which holds if the penalty $g_\lambda$ is a KL function. This is the case of all the considered hard-thresholding, soft-thresholding, SCAD and $\ell_q$-norm (with rational $q$) penalties. With the above lemmas, the proof of (45) follows similarly the proof of Theorem 1 in [30] with some minor changes, thus is omitted here for succinctness.

## APPENDIX B

The derivation of Theorem 2 follows similarly to [21, 22]. Consider the set

$$B = \left\{ \mathbf{\Delta} : \mathbf{\Delta} = \mathbf{\Delta}^T, \mathbf{\Delta} + \mathbf{\Theta}^0 \succeq \varepsilon_1 \mathbf{I}_d, \|\mathbf{\Delta}\|_F = C_1 \sqrt{\frac{s \log d}{n}} \right\}$$

where $\varepsilon_1 \leq \xi_{\min}$. Let

$$f(\mathbf{\Theta}^0 + \mathbf{\Delta}) = \frac{1}{2} \|\mathbf{\Theta}^0 + \mathbf{\Delta} - \mathbf{S}\| + g_\lambda(\mathbf{\Theta}^0 + \mathbf{\Delta})$$

if we can show that

$$\inf\{f(\mathbf{\Theta}^0 + \mathbf{\Delta}) : \mathbf{\Delta} \in B\} > f(\mathbf{\Theta}^0) \tag{46}$$

there exists at least one local minimizer $\hat{\mathbf{\Theta}}$ such that

$$\|\hat{\mathbf{\Theta}} - \mathbf{\Theta}^0\|_F \leq C_1 \sqrt{\frac{s \log d}{n}}. \tag{47}$$

Define $G(\mathbf{\Delta}) = f(\mathbf{\Theta}^0 + \mathbf{\Delta}) - f(\mathbf{\Theta}^0)$, to show (46), it is sufficient to show that $G(\mathbf{\Delta}) > 0$ for $\mathbf{\Delta} \in B$. $G(\mathbf{\Delta})$ can be expressed as

$$G(\mathbf{\Delta}) = \frac{1}{2} \|\mathbf{\Theta}^0 + \mathbf{\Delta} - \mathbf{S}\|_F^2 - \frac{1}{2} \|\mathbf{\Theta}^0 - \mathbf{S}\|_F^2 + g_\lambda(\mathbf{\Theta}^0 + \mathbf{\Delta}) - g_\lambda(\mathbf{\Theta}^0)$$
$$= \frac{1}{2} \|\mathbf{\Delta}\|_F^2 - \langle \mathbf{S} - \mathbf{\Theta}^0, \mathbf{\Delta} \rangle + g_\lambda(\mathbf{\Theta}^0 + \mathbf{\Delta}) - g_\lambda(\mathbf{\Theta}^0) \tag{48}$$

For $\sqrt{\log d / n} = o(1)$, with probability tending to 1 we have [9]

$$\max_{i \neq j} |(\mathbf{S} - \mathbf{\Theta}^0)_{ij}| \leq C_2 \sqrt{\frac{\log d}{n}}.$$

Then, it follows that

$$\langle \mathbf{S} - \mathbf{\Theta}^0, \mathbf{\Delta} \rangle \leq \left| \sum_{i \neq j} (\mathbf{S} - \mathbf{\Theta}^0)_{ij} \mathbf{\Delta}_{ij} \right| + \left| \sum_i (\mathbf{S} - \mathbf{\Theta}^0)_{ii} \mathbf{\Delta}_{ii} \right|$$
$$\leq \max_{i \neq j} |(\mathbf{S} - \mathbf{\Theta}^0)_{ij}| \|\mathbf{\Delta}\|_{1,\text{off}} \tag{49}$$
$$\leq C_2 \sqrt{\frac{\log d}{n}} \|\mathbf{\Delta}\|_{1,\text{off}}$$

with probability tending to 1, where $\mathbf{S}_{ii} = \mathbf{\Theta}_{ii}^0 = 1$ is used. Let $\mathbf{M}_S$ and $\mathbf{M}_{S^\perp}$ denote the projection of the matrix $\mathbf{M}$ to the subspace $S$ and its complement, respectively. Under the assumption that the penalty function $g_\lambda$ is *decomposable*, i.e., $g_\lambda(\mathbf{M}) = g_\lambda(\mathbf{M}_S) + g_\lambda(\mathbf{M}_{S^\perp})$, and with $\mathbf{\Theta}_{S^\perp}^0 = \mathbf{0}$, we have

$$\begin{aligned} & g_\lambda(\mathbf{\Theta}^0 + \mathbf{\Delta}) - g_\lambda(\mathbf{\Theta}^0) \\ & = g_\lambda(\mathbf{\Theta}_S^0 + \mathbf{\Delta}_S) + g_\lambda(\mathbf{\Delta}_{S^\perp}) - g_\lambda(\mathbf{\Theta}_S^0). \end{aligned} \tag{50}$$

Plugging (49) and (50) into (48) yields

$$G(\mathbf{\Delta}) \geq \frac{1}{2} C_1^2 \frac{s \log d}{n} \underbrace{- C_2 \sqrt{\frac{\log d}{n}} \|\mathbf{\Delta}_S\|_1}_{K_1} + \underbrace{g_\lambda(\mathbf{\Theta}_S^0 + \mathbf{\Delta}_S) - g_\lambda(\mathbf{\Theta}_S^0)}_{K_2}$$
$$+ \underbrace{g_\lambda(\mathbf{\Delta}_{S^\perp}) - C_2 \sqrt{\frac{\log d}{n}} \|\mathbf{\Delta}_{S^\perp}\|_{1,\text{off}}}_{K_3}. \tag{51}$$

Next, we bound the three terms $K_1$, $K_2$ and $K_3$. Since $\|\mathbf{\Delta}_S\|_1 \leq \sqrt{s} \|\mathbf{\Delta}_S\|_F \leq \sqrt{s} \|\mathbf{\Delta}\|_F$, we have

$$K_1 \leq C_1 C_2 \frac{s \log d}{n}. \tag{52}$$

Using Taylor's expansion of $f(t) = g_\lambda(\mathbf{\Theta}_S^0 + t \mathbf{\Delta}_S)$ with integral form of the remainder, it follows that

$$|K_2| \leq \left|\langle g'_\lambda(\boldsymbol{\Theta}^0_S), \boldsymbol{\Delta}_S \rangle\right|$$
$$+ \left|\text{vec}(\boldsymbol{\Delta}_S)^T \left[\int_0^1 (1-t) g''_\lambda(\boldsymbol{\Theta}^0_S + t\boldsymbol{\Delta}_S) dt\right] \text{vec}(\boldsymbol{\Delta}_S)\right|$$
$$\leq \max_{(i,j)\in S} \left|g'_\lambda(\boldsymbol{\Theta}^0_{ij})\right| \|\boldsymbol{\Delta}_S\|_1 + o(1)\|\boldsymbol{\Delta}_S\|_F^2 \quad (53)$$
$$= O\left(C_1 \frac{s \log d}{n}\right) + o(1) C_1^2 \frac{s \log d}{n}$$

where the conditions $\max_{(i,j)\in S} g'_\lambda(\boldsymbol{\Theta}^0_{ij}) = O(\sqrt{\log d/n})$ and $\max_{(i,j)\in S} g''_\lambda(\boldsymbol{\Theta}^0_{ij}) = o(1)$ are used. Moreover, since

$$\sup_{\|\boldsymbol{\Delta}\|_F \in \mathbb{B}} \boldsymbol{\Delta}_{ij} \leq C_1 \sqrt{\frac{s \log d}{n}} = O(\lambda)$$

we can find a constant $C_3 > 0$ such that $g_\lambda(\boldsymbol{\Delta}_{ij}) \geq \lambda C_3 |\boldsymbol{\Delta}_{ij}|$ for all $(i,j) \in S^\perp$. Then, it follows that

$$K_3 \geq \left(\lambda C_3 - C_2 \sqrt{\frac{\log d}{n}}\right) \|\boldsymbol{\Delta}_{S^\perp}\|_{1,\text{off}} \geq 0 \quad (54)$$

since $\lambda = O(\sqrt{s\log d/n})$. Consequently, by taking a sufficiently large constant $C_1$, we have $G(\boldsymbol{\Delta}) \geq 0$ with probability tending to 1, which complete the proof. For the $\ell_1$-norm penalty, we can set $\lambda = C_3 \sqrt{\log d/n}$ with $C_3 > C_2$ such that (10) holds, since $g_\lambda(\boldsymbol{\Delta}_{S^\perp}) = \lambda \|\boldsymbol{\Delta}_{S^\perp}\|_{1,\text{off}}$.

The proof of the spectral norm result follows from the same argument as Theorem 4.2 in [20], which is omitted for conciseness.

## APPENDIX C

The objective function in the $\boldsymbol{\Sigma}$-subproblem (28) is quadratic in $\boldsymbol{\Sigma}$, thus, it has an analytical solution. Specifically, from the first-order optimality condition, the minimizer of the $\boldsymbol{\Sigma}$-subproblem satisfies

$$\mathbf{A}^T \mathbf{A} \boldsymbol{\Sigma}^{k+1} \mathbf{A}^T \mathbf{A} + 2\rho \boldsymbol{\Sigma}^{k+1} = \mathbf{A}^T \mathbf{Y} \mathbf{A} + \rho \mathbf{G}^T \boldsymbol{\Gamma}^{k+1}. \quad (55)$$

We can construct a matrix $\boldsymbol{\Sigma}^{k+1}$ that satisfies this condition and thus minimizes the objective. Let $\mathbf{Z}^k = \mathbf{A}^T \mathbf{Y} \mathbf{A} + \rho \mathbf{G}^T \boldsymbol{\Gamma}^{k+1}$, it follows that

$$\boldsymbol{\Lambda} \mathbf{E}^T \boldsymbol{\Sigma}^{k+1} \mathbf{E} \boldsymbol{\Lambda} + 2\rho \mathbf{E}^T \boldsymbol{\Sigma}^{k+1} \mathbf{E} = \mathbf{E}^T \mathbf{Z}^k \mathbf{E}$$

which can be equivalently written as

$$(\mathbf{a}\mathbf{a}^T) \circ (\mathbf{E}^T \boldsymbol{\Sigma}^{k+1} \mathbf{E}) + (2\rho \mathbf{1}_{n\times n}) \circ (\mathbf{E}^T \boldsymbol{\Sigma}^{k+1} \mathbf{E}) = \mathbf{E}^T \mathbf{Z}^k \mathbf{E}. \quad (56)$$

Then, it follows from (56) that

$$\mathbf{E}^T \boldsymbol{\Sigma}^{k+1} \mathbf{E} = [\mathbf{E}^T \mathbf{Z}^k \mathbf{E}] \oslash (\mathbf{a}\mathbf{a}^T + 2\rho \mathbf{1}_{n\times n}) \quad (57)$$

which finally results in (31).